# Negligible current-induced torque from the bulk and interface of Al

Taiyang Zhang,[1,2] Lujun Zhu,[3] Zhihao Yan,[1,2] Lijun Zhu[1,2]*
[1] *State Key Laboratory of Semiconductor Physics and Chip Technologies, Institute of Semiconductors, Chinese Academy of Sciences, Beijing 100083, China*
[2]*Center of Materials Science and Optoelectronics Engineering, University of Chinese Academy of Sciences, Beijing 100049, China*
[3]*College of Physics and Information Technology, Shaanxi Normal University, Xi'an 710062, China*
**ljzhu@semi.ac.cn*

The light metal Al was predicted to have strong orbital Hall and Rashba effects from its bulk and interface, respectively. In this letter, we report experimental evidence that neither the bulk nor the interface of the Al contributed a detectable torque on adjacent Co and FePt layers with significant spin Hall effect, spin-orbit coupling, and resistivity mismatch with Al. These results suggest minimal orbital-spin conversion in Al and negligible orbital transport and spin-vorticity torque in Co/Al, Al/Co, and Al/FePt bilayers. Our findings suggest poor generality and/or effectiveness of torque contributions by the orbital Hall effect, the interfacial orbital Rashba effect, and the spin-vorticity coupling.

In the past two decades, spin current from the spin Hall effect (SHE) is consensually known to exert strong spin-orbit torques (SOTs) on nanomagnets via exchange interaction of spin polarization and the magnetization [1-7], while orbital polarization cannot directly interact with magnetization (Fig. 1a)[8]. Recently, there is a heated debate as to whether orbital current, i.e., flow of orbital polarization, can be a new effective source of spin current for torque generation [9-22]. So far, orbital polarization has been claimed from the orbital Hall conductivity [9-14], the interfacial orbital Rashba effect [15-18], the orbital vorticity at magnetic interfaces with resistivity gradients [21,22]. Meanwhile, there have been several mechanisms claimed for orbital-spin conversion, e.g., the spin-orbit coupling (SOC) of a bulk or interface, the spin-orbit polarization (a parameter depending on the SOC and band filling)[19,20], spin-vorticity coupling [21,22], etc.

In sharp contrast, some other theories suggest ultrashort orbital relaxation length regardless of the SOC strength [8], which further implies minimal intralayer orbital accumulation and interlayer transport of orbital polarization. Meanwhile, orbital-spin conversion was also predicted to be possible only in spin Hall materials [8], suggesting little role of orbital polarization in weak SHE materials. These essential disagreements require urgent clarification of the general effectiveness of orbital contribution to the SOTs in normal metal/ferromagnet (FM) heterostructures. Previously, Al was predicted to have negligible SHE, high orbital Hall conductivity of -700 ($\hbar$/e) $\Omega^{-1}$ $cm^{-1}$ [9], and significant orbital Rashba effect at the Co/Al interface [16,17], providing an ideal platform for clarifying the general effectiveness of orbital torque contributions.

In this letter, we investigate the torque effect from the predicted orbital polarization in the light metal Al. Our experiments reveal that that neither the bulk nor the interface of the Al contributed a detectable torque on adjacent magnetic layer with strong SHE and SOC. These results suggest minimal orbital-spin conversion in Al and negligible orbital transport and spin-vorticity torque in Al/FM bilayers, regardless of the SOC of the FM.

*Devices and characterizations.* Devices for this work includes Hall bars of Pt 5/Co 2/Al 0-4 trilayers, FePt 4/Al 0-4 bilayers, and Al 1-3.1/Co 2 bilayers (numbers are the layer thicknesses in nm). The 2 nm Co and 4 nm FePt are chosen as the FMs for their negligible self-induced SOTs [23-25], significant SHE [26-28], and large resistivity mismatch with Al (see below). The Pt layer in the Pt/Co/Al trilayers is to make the Co smooth and to demonstrate the high sensitivity of our torque measurements. We do not use $SiO_2$/Co/Al samples for the SOT study because Co grown on oxidized Si substrates is strongly granular and too rough to be used for the SOT study [29]. Each Hall bar device is sputter-deposited on thermally oxidized Si substrates at room temperature under background vacuum of $<10^{-9}$ Torr and protected by a MgO 1.6/Ta 1.5 bilayer that is fully oxidized upon exposure to the atmosphere [30]. A 1 nm amorphous Ta was deposited as the adhesion layer for the Pt/Co/Al samples but not for the FePt/Al and Al/Co samples. The cross-sectional transmission electron microscopy imaging reveals reasonably smooth interfaces for the Pt/Co/Al and FePt/Al samples (Fig. 1b). The dimensions of Hall bars are defined as 5 μm × 60 μm by photolithography and argon ion milling (Fig. 1c). The in-plane magnetization hysteresis (Fig. 1d) and the planar Hall voltage (Fig. 1e) measurements reveal in-plane magnetic anisotropy and saturation magnetization ($M_s$) of 1255 emu/$cm^3$ for the Co and 837 emu/$cm^3$ for the FePt, which are consistent with our previous reports [24,25].

*Spin-orbit torque characterizations.* The dampinglike and fieldlike SOTs of the Pt/Co/Al, the FePt/Al, and the Al/Co are quantified from angle-dependent harmonic Hall voltage measurements with the thermal effects being carefully taken into account [7,23,31-34]. While a sinusoidal electric field ($E$) is applied onto the Hall bar (Fig. 1c), the first and second harmonic Hall voltages ($V_{1\omega}$ and $V_{2\omega}$) are collected as a function of the angle ($\varphi$) of the in-plane magnetic field ($H_{xy}$) relative to $E$. $E$ is about 19 kV/m for the Pt/Co/Al devices and 35 kV/m for the more resistive FePt/Al and Al/Co devices. The magnitude of $H_{xy}$ is varied between 0.75 kOe and 3.75 kOe to assure the macrospin behavior (Fig. 1d)[33] and negligible ordinary Nernst effect [34]. When an IMA macrospin sample interacts with a transversely polarized spin current, $V_{2\omega}$ reads [32]

$$V_{2\omega} = V_{\mathrm{DL+ANE}}\cos\varphi + V_{\mathrm{FL+Oe}}\cos\varphi\cos2\varphi + V_{\mathrm{PNE}}\sin2\varphi, \quad (1)$$

with

$$V_{\mathrm{DL+ANE}} = V_{\mathrm{AHE}}H_{\mathrm{DL}}/2(H_{xy}-H_{\mathrm{k}}) + V_{\mathrm{ANE}}, \quad (2)$$

$$V_{\mathrm{FL+Oe}} = -V_{\mathrm{PHE}}(H_{\mathrm{FL}}+H_{\mathrm{Oe}})/H_{xy}. \quad (3)$$

Here, $H_{\mathrm{DL}}$ is the dampinglike SOT field, $H_{\mathrm{FL}}$ is the fieldlike SOT field, $H_{\mathrm{Oe}}$ is the transverse Oersted field exerted on the magnetic layer by the in-plane charge current in other layers, $H_k$ is the effective PMA field (negative for IMA samples), $V_{AHE}$ is the anomalous Hall voltage, $V_{ANE}$ is the anomalous Nernst voltage induced by the vertical thermal gradient, $V_{PNE}$ is the planar Nernst voltage induced by the longitudinal thermal gradient [32], and $V_{PHE}$ is the planar Hall voltage ($V_{1\omega} = V_{PHE}\sin2\varphi$ under zero $H_z$, Fig. 1e). The detailed values of $H_k$, $V_{AHE}$, $V_{ANE}$, $V_{PHE}$, and $E$ are listed in Table S1 in the Supplementary Materials [35].

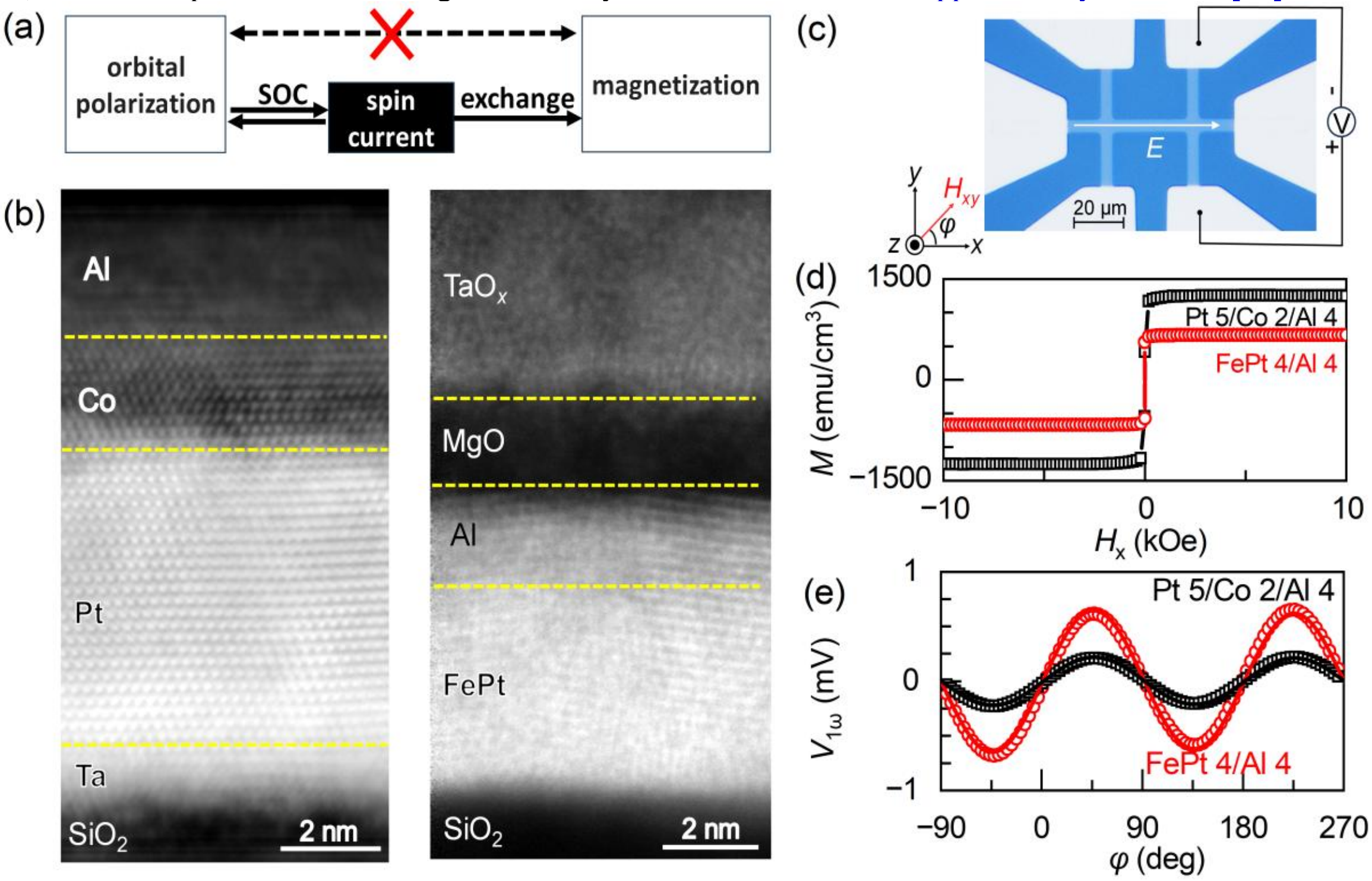

**Figure 1** (a) Schematic of potential path for orbital torque generation, highlighting that orbital polarization cannot interact with magnetization unless being converted into spin current via SOC. (b) Cross-sectional transmission electron microscopy images of a Pt/Co/Al sample and a FePt/Al sample. (c) Optical microscopy image of the Hall bar device and the measurement coordinate. (d) Magnetization hysteresis and (e) First harmonic Hall voltage vs the azimuth angle ($\varphi$) of the in-plane magnetic field ($H_{xy}$ = 3.5 kOe) for the Pt 5/Co 2/Al 4 and the FePt 4/Al 4.

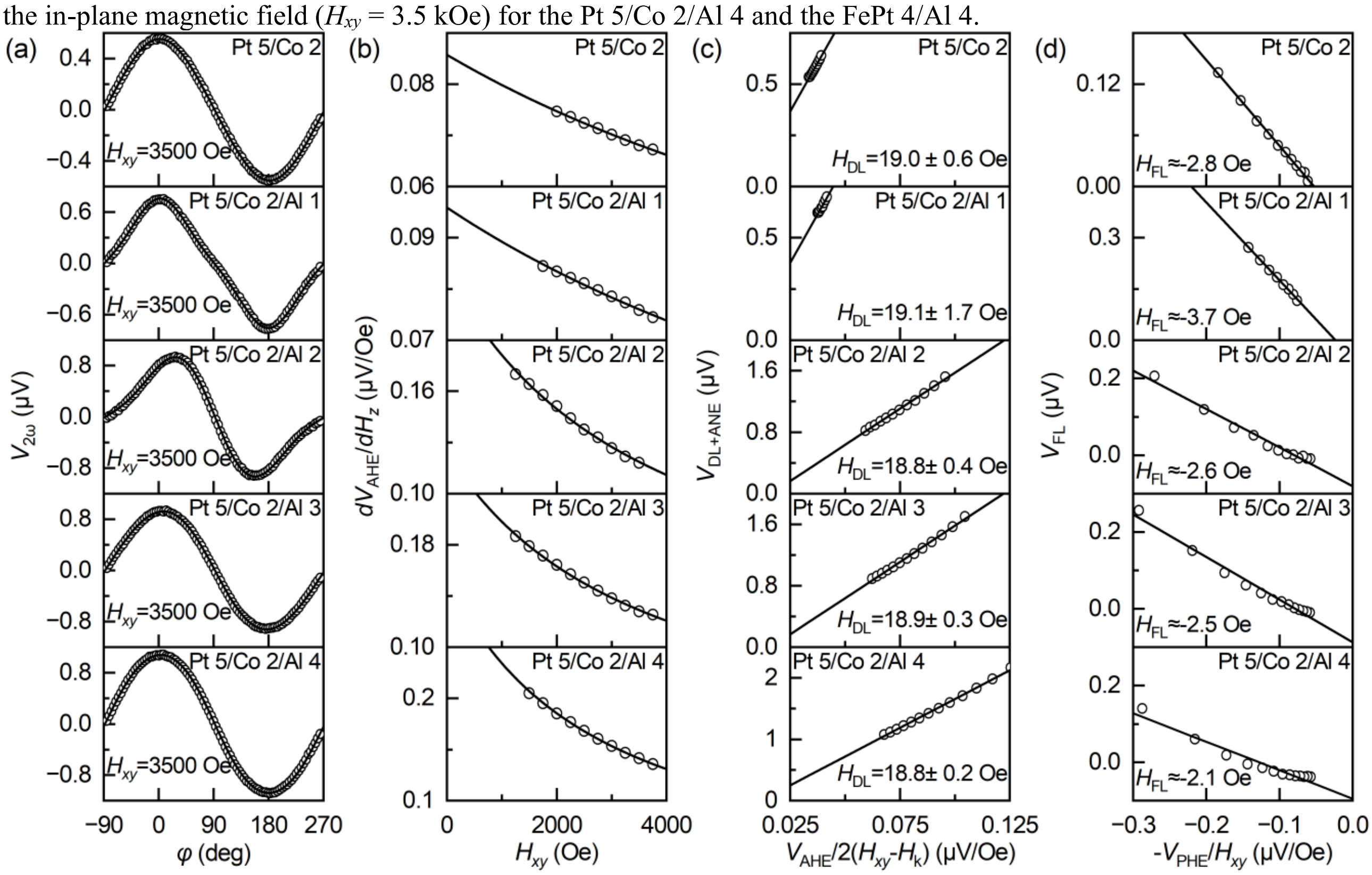

Figure 2. Harmonic Hall voltage analysis of the Pt/Co/Al devices. (a) $V_{2\omega}$ vs $\varphi$ ($H_{xy}$ = 3.5 kOe), (b) $dV_{AHE}/dH_z$ vs $H_{xy}$, (c) $V_{DL+ANE}$ vs $V_{AHE}/2(H_{xy}$-$H_k)$, and (d) $V_{FL+Oe}$ vs -$V_{PHE}/H_{xy}$ for the FePt/Al bilayers with Al thickness of 0 nm, 1 nm, 2 nm, 3nm, and 4 nm, respectively. The solid lines in (a)-(d) represent the best fits of the data to Eq. (1), Eq. (4), Eq. (2), and Eq. (3), respectively. The electric field $E$ is about 19 kV/m.

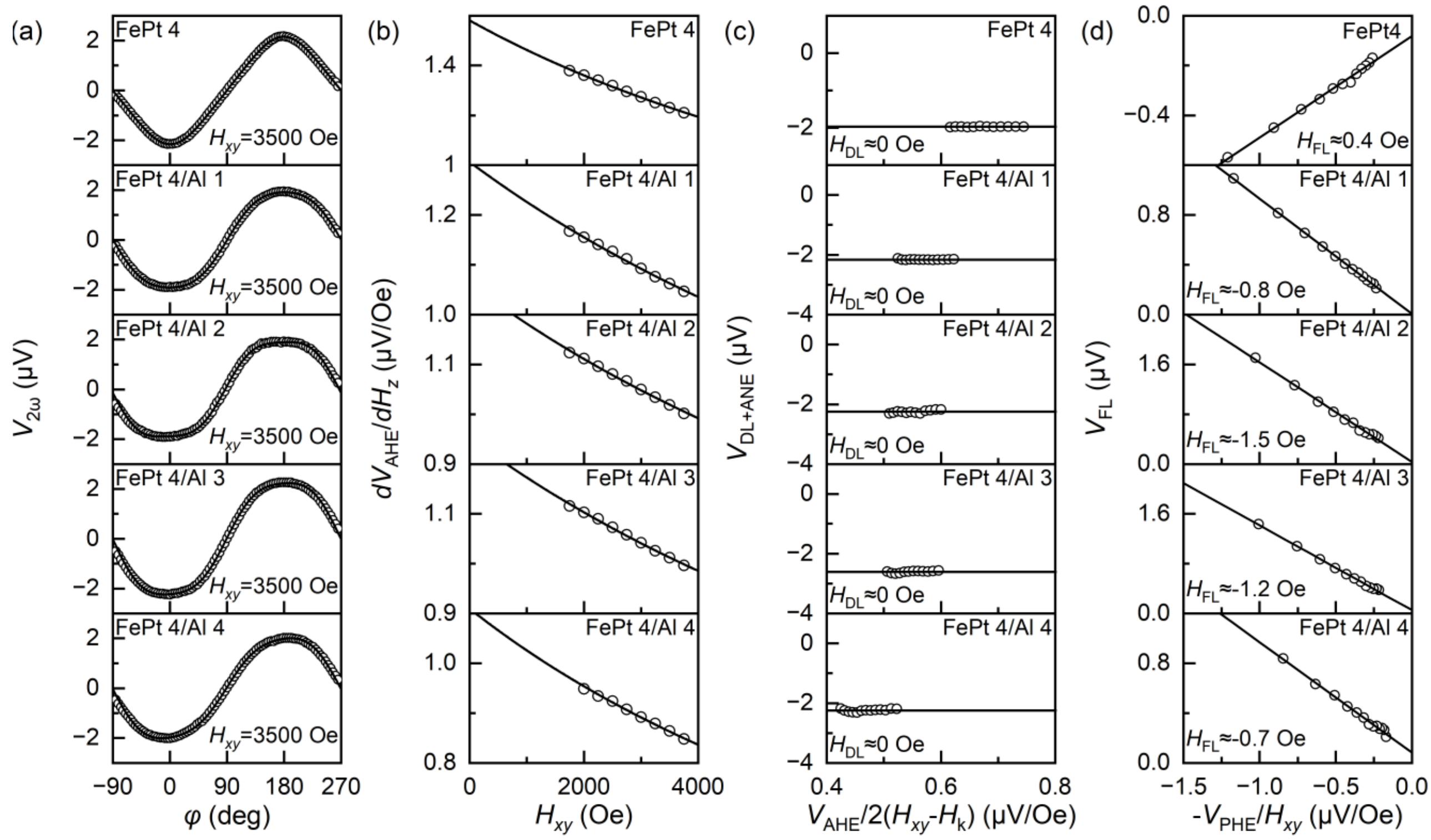

Fig. 3. Harmonic Hall voltage analysis of the FePt/Al bilayers. (a) $V_{2\omega}$ vs $\varphi$ ($H_{xy}$ = 3.5 kOe), (b) $dV_{AHE}/dH_z$ vs $H_{xy}$, (c) $V_{DL+ANE}$ vs $V_{AHE}/2(H_{xy}-H_k)$, and (d) $V_{FL+Oe}$ vs $-V_{PHE}/H_{xy}$ for the FePt/Al bilayers with Al thickness of 0 nm, 1 nm, 2 nm, 3nm, and 4 nm, respectively. The solid lines in (a)-(d) represent the best fits of the data to Eq. (1), Eq. (4), Eq. (2), and Eq. (3), respectively. The electric field $E$ is about 35 kV/m.

For clarity, we show the harmonic Hall voltage measurement data of the Pt/Co/Al devices in Fig. 2a-d and the FePt/Al devices in Fig. 3a-d. The values of $V_{DL+ANE}$ and $V_{FL+Oe}$ for each magnitude of $H_{xy}$ are determined from the fits of $V_{2\omega}$ vs $\varphi$ to Eq. (1) (Fig. 2a and Fig.3a). As shown in Fig. 2b and Fig.3b, the values of $H_k$ and $V_{AHE}$ are determined from the fits of $dV_{1\omega}/dH_z$ vs $H_{xy}$ to [24]

$$dV_{1\omega}/dH_z = V_{AHE}/(H_{xy}-H_k), \tag{4}$$

where $dV_{1\omega}/dH_z$ is measured from the linear scaling of $V_{AHE}$ with the swept out-of-plane field ($H_z$) under given in-plane magnetic field $H_x$. As shown in Fig. 2c,d and Fig. 3c,d, the values of $H_{DL}$ and $H_{FL}$ are estimated from the slopes of the linear fits of $V_{DL+ANE}$ vs $V_{AHE}/2(H_{xy}-H_k)$ to Eq. (2) and of $V_{FL+Oe}$ vs $-V_{PHE}/H_{xy}$ to Eq. (3), respectively. $H_{Oe}$ is estimated as $(j_{Pt}d_{Pt}-j_{Al}d_{Al})/2$ for the Pt/Co/Al devices and as $-j_{Al}d_{Al}/2$ for the FePt/Al devices. Here, $j_{Pt(Al)} = E/\rho_{Pt(Al)}$ and $d_{Pt}(d_{Al})$ are the current density and the thickness of the Pt (Al) layers. The resistivity of the Pt layer ($\rho_{Pt}$) is 34 μΩ cm as determined from a control Pt device without Co and Al layers. The resistivity of the Al layers ($\rho_{Al}$) within the Pt/Co/Al (FePt/Al) devices are 126-190 μΩ cm (414-664 μΩ cm) as determined from the conductance difference between the devices with $d_{Al}$ =1-4 nm and with $d_{Al}$ =0 nm. With the values of $H_{DL}$ and $H_{FL}$ of the Pt/Co/Al and the FePt/Al, the dampinglike and fieldlike SOT efficiencies per electric field ($\xi_{DL}^{E}$ and $\xi_{FL}^{E}$) are estimated following

$$\xi_{\mathrm{DL}(FL)}^{E} = (2e/\hbar)\mu_0 M_s t_{FM} H_{\mathrm{DL(FL)}}/E, \tag{5}$$

where $e$ is the elementary charge, $\hbar$ the reduced Planck's constant, $\mu_0$ the permeability of vacuum, and $t_{FM}$ the total thickness of the magnetic layer.

As plotted in Fig. 4, $\xi_{DL}^{E}$ for the Pt/Co/Al remains essentially at the same high value of $(7.66\pm0.09)\times10^5$ $\Omega^{-1}$ $m^{-1}$ as the Al thickness ($d_{Al}$) is varied from 0 nm to 4 nm. In another word, whether or not there was Al and Co/Al interface, $\xi_{DL}^{E}$ of the Pt/Co remains unchanged. This result reveals negligible torque contribution from the theoretically predicted bulk orbital Hall conductivity in the Al layer and the orbital Rashba effect of the Co/Al interface. The fieldlike torque of $\xi_{FL}^{E}$= $(1.11 \pm 0.23)\times10^5$ $\Omega^{-1}$ $m^{-1}$ is much weaker than $\xi_{DL}^{E}$ and of the opposite sign compared to the dampinglike torque. This observation is consistent with the SHE of the Pt being the only origin of the observed dampinglike and fieldlike torques [7,36]. The very slight variation of the fieldlike torque seems to arise from the uncertainty of sample fabrications rather than from any Rashba effect at the Co/Al interface because the interfacial Rashba torque should vary more significantly at very small thicknesses than at large thicknesses, without changing sign [37]. For the FePt, both $\xi_{DL}^{E}$ and $\xi_{FL}^{E}$ are negligibly small for all the Al thicknesses (0-4 nm). The FePt/Al results reveal that the theoretically-predicted orbital Hall effect of the Al generates no torques on the adjacent FePt layer even though the bulk SOC of the magnetic layer is very strong. The negligible self-induced SOT within the 4 nm FePt is also consistent with our previous reports [24,25]. The negligible contribution of the Al on the SOTs is consistent with the previous reports of negligible torque contributions of the theoretically calculated orbital Hall conductivity in Ta and Ti to their heterostructures [23,24,38-40]. As plotted in Fig. 4, both $\xi_{DL}^{E}$ and $\xi_{FL}^{E}$ of the control Al/Co devices are also negligibly small (see Fig. S2 for details of the harmonic Hall analysis [35]), reaffirming negligible torque contribution from the bulk and interface of the Al.

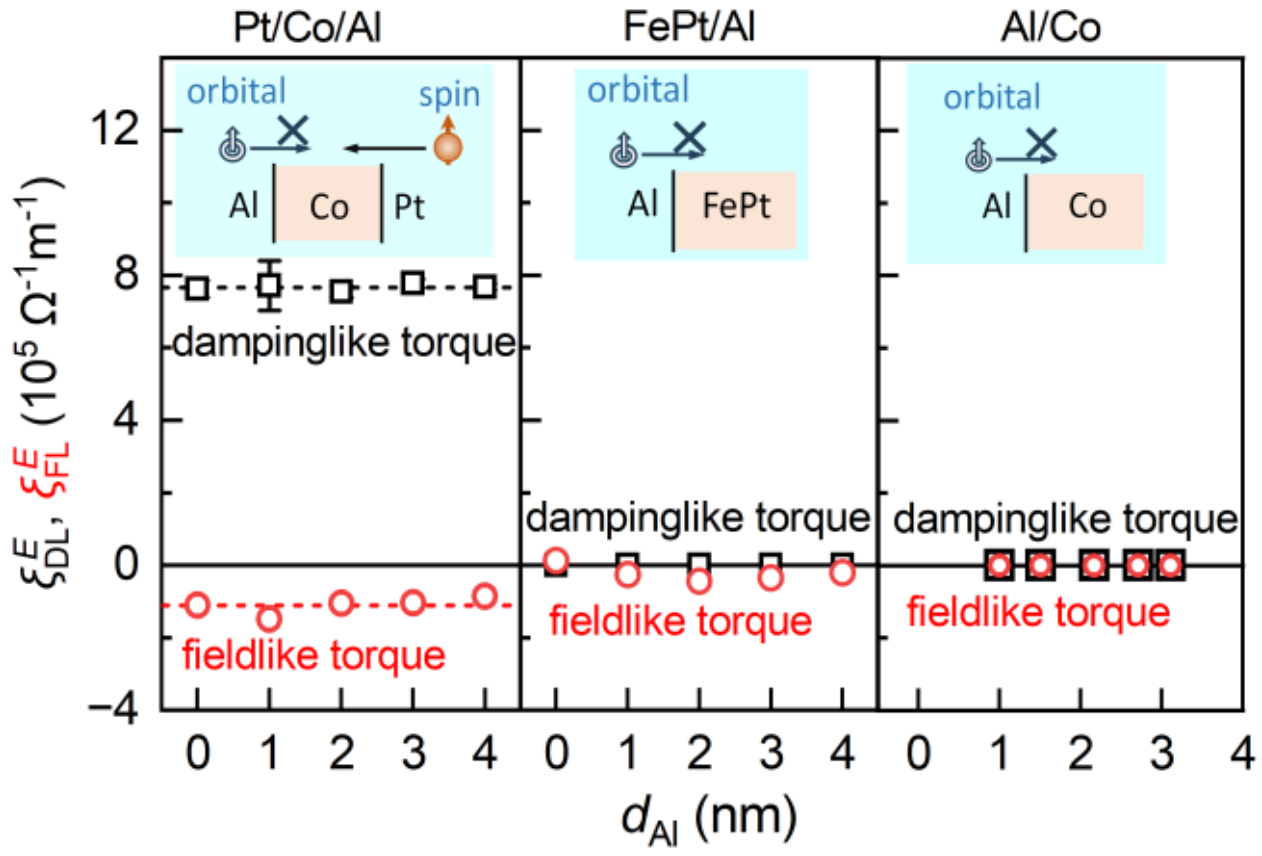

Fig. 4. Dampinglike and fieldlike SOT efficiencies per electric field for the Pt/Co/Al, the FePt/Al, and the Al/Co with varying Al thickness. The dashed line is to guide the eyes. Error bars are standard deviations. Insets suggest the orbital and spin transport into the magnetic layer.

*Discussion.* Microscopically, the negligible SOT contributions from Al and Co/Al interface in the Pt/Co/Al, the FePt/Al, and the Al/Co must indicate little orbital-spin conversion in Al and negligible injection of orbital polarization into the magnetic layers (Co and FePt) from the Al and its interface (see the insets of Fig. 4). A significant spin current from orbital-spin conversion in Al must be detected by the FMs, while orbital polarization injected to Co or FePt would be partially converted into spin current for torque generation. Note that Co and FePt have significant SHE and bulk SOC (84 meV for Co [41] and 335 meV for FePt [25]) necessary for orbital-spin conversion. In addition, the Co/Al also has significant interfacial SOC at the Co/Al interface. As indicated by the Bruno's model [42] and previous experiments [43,44], the SOC of a magnetic interface can be parameterized using the PMA energy density ($K_s$) of the interface. Following the relation of $H_k$ = -4π$M_s$ + 2$K_s$/$M_s t_{Co}$, $K_s$ of the two Co interfaces is estimated to be 0.28±0.04 erg/cm$^2$ for the Pt/Co/MgO, be 0.30±0.06 erg/cm$^2$ for Pt/Co/Al 1 to 1.55±0.07 in average for Pt/Co/Al 2-4 (see more details in Fig. S1 in the Supplementary Material [35]). Assuming an approximately constant PMA for the bottom Pt/Co interface of the Pt/Co/Al samples, $K_s$ for the Co/Al interface would be greater than ≈1.2 erg/cm$^2$, which is giant compared to that of typical magnetic interfaces [44].

The observation of negligible orbital-spin conversion in Al and interlayer orbital transport is consistent with the predictions of ultrashort orbital relaxation length of Al (regardless of the SOC) and orbital-spin conversion by the SHE in the recent theory [8]. Note that orbital quenching due to strong crystal field of Al and rapid orbital relaxation at interfaces are generally expected [8]. As discussed by another recent theory [22], one can even doubt about the reliability of the first-principles or tight-band calculations of the orbital Hall conductivity and orbital Rashba effect. Despite the previous report in Cu/PtCo [21], the spin-vorticity torque appears ineffective in Co/Al, Al/FePt, and Al/Co devices with significant resistivity mismatch (e.g., 63 μΩ cm for the Co, 126-190 μΩ cm for the Al in the Pt/Co/Al, 110 μΩ cm for the FePt, 414-664 μΩ cm for the Al in FePt/Al). The absence of detectable spin-vorticity torque is consistent with previous experiments on Ti/FM and Cu/FM bilayers [23,24,38-40].

Finally, our conclusions are not affected by a single field-scan harmonic Hall experiment [45] claiming that a 1-5 nm Al layer within some Pt 8/Co 0.9/Al/Pt 3 samples increased the dampinglike and fieldlike SOTs by up to 5.5 and 14 times. That analysis assumed a continuous tilting of a perpendicular macrospin from the film normal to the film plane by sweeping the magnetic field in the large scale along the current direction. As we have discussed previously [46-48], the macrospin assumption becomes inapplicable when a perpendicular magnetization is tilted far away from the easy axis. Instead, an in-plane magnetic field can usually drive a PMA sample into multi-domain configurations or even sharply switch it [48].

*Conclusion.* We have demonstrated that neither the bulk nor the interface of the Al contributed a detectable torque on adjacent magnetic layer with strong SHE and SOC, despite the theoretically predicted bulk orbital Hall effect, interfacial orbital Rashba effect, and resistivity difference between Al and the ferromagnets Co and FePt. These results suggest minimal orbital-spin conversion in Al and negligible interlayer orbital transport and spin-vorticity torque in Al/FM and FM/Al bilayers. The absence of distinguishable orbital torque from Al, together with that from Ti [23,24,38-40], Ta [25], Cu [23], and W [49], $CuO_x$ [50], consistently reveals the poor generality and/or effectiveness of torque contributions by the orbital Hall effect, interfacial orbital Rashba effect, and spin-vorticity coupling. Our findings are highly stimulating for further clarifying the generality and effectiveness of orbital torque contributions utilizing other material systems.

*Acknowledgement.* The authors thank Albert Fert, Ke Xia, Xiangrong Wang, Kai Chang, Jianhua Zhao, and Hongxing Yang for helpful discussions. This work is supported partly by the Beijing Natural Science Foundation (Z230006), the National Natural Science Foundation of China (12274405), and the Natural Science Foundation of Shaanxi Province (2024JC-YBMS-315).